# Electron capture beta decay of partially polarized nuclei

R.Basak and V. I. Tsifrinovich

Department of Applied Physics, NYU Tandon School of Engineering, Brooklyn, NY 11201

## Abstract

We compute the spin excess for the neutrinos radiated in the process of electron capture beta decay of partially polarized nuclei. The results of computation are presented for the $^{119}$Sb nuclei polarized by the strong hyperfine field in a ferromagnetic substance. This system was suggested as a possible source of directed neutrino radiation. We directly compute the spin excess of radiated neutrinos and show that it is greater than that estimated previously under simplifying assumptions.

## I. Introduction

Electron capture beta decay of nuclei polarized by a strong hyperfine field in a ferromagnetic substance can be used as a compact source of directed mono-energetic neutrino radiation [1]. Such a device could be exploited in future neutrino communication [2,3], for the measurement of the neutrino rest mass [4], and for testing of Lorentz invariance in beta decay [5]. Probably, the most appropriate nuclei for these objectives are $^{119}$Sb with the nuclear spin $I$ = 5/2, nuclear magnetic moment μ = 1.74x10$^{-26}$ J/T and half-life $T_{1/2}$ = 38.2 h[6-8]. The hyperfine field on the Sb nuclei may be as high as 70.6 T [9].

The decay scheme for the primary K-capture branch of [119]Sb is simple: a nucleus absorbs a 1S-electron and decays into the ground state of [119]Sn ($I$ = 3/2) with emission of a 538 keV neutrino. If the temperature of a magnetized sample containing [119]Sb atoms is well below the hyperfine splitting then the parent nuclei occupy the lowest hyperfine level $I_z = m = 5/2$ (we assume that the hyperfine field points "up", i.e. in the positive z-direction). Due to the conservation of the z-component of the total angular momentum both the remaining electron spin and the neutrino spin must point up. Finally, due to the helicity requirements the neutrino generated in this reaction will propagate in the negative z-direction.

## II. Calculations

At temperatures comparable with the hyperfine splitting the parent nuclei occupy all the hyperfine levels. In this case the important parameter describing the neutrino radiation is the spin excess, i.e. the difference between the relative probabilities of radiating neutrinos with "up" and "down" spins. In this paper we compute the neutrino spin excess $S_\nu$:

$$S_\nu = \sum_{m=-I}^{I} P_m \frac{P_{m\uparrow} - P_{m\downarrow}}{P_{m\uparrow} + P_{m\downarrow}} \qquad (1)$$

Here $P_m$ is the probability to occupy the parent nuclear Zeeman level $m$, $P_{m\uparrow}$ is the probability of decay from level $m$ with spin "up", and $P_{m\downarrow}$ is the same for spin "down". The probability to occupy level $m$ is given by the Boltzmann's factor:

$$P_m = \left( \sum_{m=-I}^{I} e^{-\frac{\mu B m/I}{k_B T}} \right)^{-1} e^{-\frac{\mu B m/I}{k_B T}}, \qquad (2)$$

where $B$ is the sum of the hyperfine field and the external magnetic field on the nuclei. In ref. [1] the spin excess was estimated assuming that the probability $P_{m\uparrow}(P_{m\downarrow})$ is proportional to the number of channels generating the neutrino with spin "up" ("down"). In this work we compute the spin excess directly without any simplifying assumptions.

The spin dependence of the decay probability for every decay channel is described by the squared modulus of the product of two Wigner3j symbols which we denote as $W$:

$$W = W(m, m', m_\nu) = \left| \begin{pmatrix} I' & K & I \\ -m' & \Delta m & m \end{pmatrix} \begin{pmatrix} S_\nu & K & S_e \\ -m_\nu & -\Delta m & m_e \end{pmatrix} \right|^2 \qquad (3)$$

(See, for example, ref. [10].) Here $I'$ and $m'$ are the spin and its z-component for the daughter nucleus, $K = I - I'$, $\Delta m = m' - m$, $m_e$ and $m_\nu$ are the spin z-components of the absorbed electron and the emitted neutrino. The first 3j symbol describes the change of the nuclear spin and the second one the change of the lepton spin. In our case the first row in the 3j symbols is fixed: $I = 5/2$, $I' = 3/2$, $K = 1$, $S_e = S_\nu = 1/2$.

As an example, for the lowest Zeeman level $m = 5/2$ the conservation of the angular momentum z-component allows only one decay channel: $m' = 3/2$, $m_e = -1/2$, and $m_\nu = 1/2$, i.e. the z-component of the angular momentum decreases by one unit for the nuclear spin and increases by the same amount for the lepton system. The corresponding value of $W$ is:

$$W\left(\frac{5}{2},\frac{3}{2},\frac{1}{2}\right) = \left| \begin{pmatrix} \frac{5}{2} & 1 & \frac{3}{2} \\ -\frac{3}{2} & -1 & \frac{5}{2} \end{pmatrix} \begin{pmatrix} \frac{1}{2} & 1 & \frac{1}{2} \\ -\frac{1}{2} & 1 & -\frac{1}{2} \end{pmatrix} \right|^2 = \frac{1}{18} \qquad (4)$$

(The sum of the elements in the second row of a 3j symbol is always equal to zero.) For the state $m = -5/2$ we also have only one decay channel, for the states $m$ ±3/2 three channels, and for the states m ±1/2 four channels. All these channels with the corresponding values of $W$ are shown in Table 1.

| m | m' | $m_\nu$ | W | m | m' | $m_\nu$ | W |
|---|---|---|---|---|---|---|---|
| 5/2 | 3/2 | 1/2 | 1/18 | -1/2 | 1/2 | -1/2 | 1/60 |
| 3/2 | 3/2 | 1/2 | 1/90 | | -1/2 | -1/2 | 1/60 |
| | 3/2 | -1/2 | 1/90 | | -1/2 | 1/2 | 1/60 |
| | 1/2 | 1/2 | 1/30 | | -3/2 | 1/2 | 1/180 |
| 1/2 | 3/2 | -1/2 | 1/180 | -3/2 | -1/2 | -1/2 | 1/30 |
| | 1/2 | -1/2 | 1/60 | | -3/2 | -1/2 | 1/90 |
| | 1/2 | 1/2 | 1/60 | | -3/2 | 1/2 | 1/90 |
| | -1/2 | 1/2 | 1/60 | -5/2 | -3/2 | -1/2 | 1/18 |

Table 1. The values of $W$ for all the decay channels.

The total decay probability from an arbitrary Zeeman level $m$ is proportional to the sum of the values of $W$ for all the decay channels from level $m$. This sum is the same for every Zeeman level:

$$\sum_{m'} W\left(m,m',\frac{1}{2}\right) + \sum_{m'} W\left(m,m',-\frac{1}{2}\right) = \frac{1}{18} \quad , \qquad (5)$$

Now the expression for the spin excess (1) can be rewritten in the form:

$$S_\nu = 18 \sum_{m=-I}^{I} P_m \left\{ \sum_{m'} W\left(m,m',\frac{1}{2}\right) - \sum_{m'} W\left(m,m',-\frac{1}{2}\right) \right\} \qquad (6)$$

### III. Calculations Results and Discussion

In Fig. 1 (solid line) we present the neutrino spin excess as a function of temperature for $^{119}$Sb for the magnetic field B = 70.6T. For comparison, with a dashed line, we present the neutrino spin excess obtained in ref. [1]. One can see that the spin excess computed in this paper is greater than that computed under simplifying assumptions.

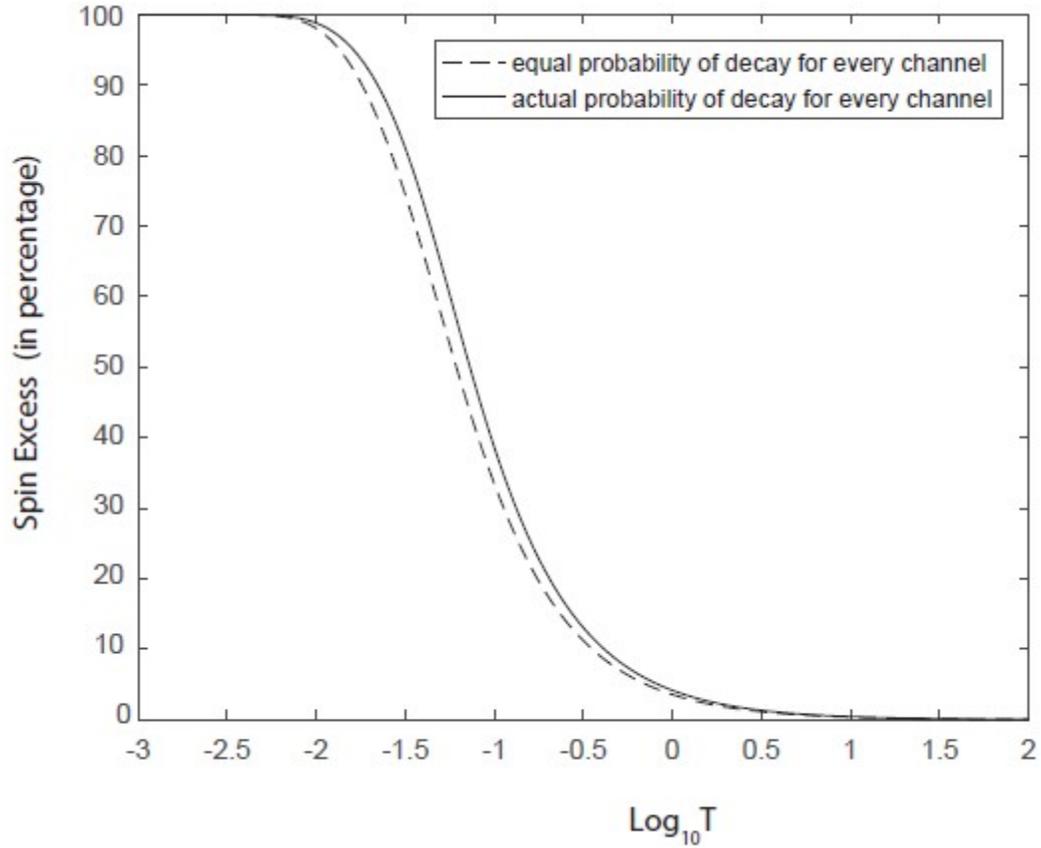

Fig. 1. Neutrino spin excess as a function of temperature (solid line). The dashed curve represents the spin excess obtained in [1].

Probably, the simplest way to detect the directed neutrino radiation is a measurement of the recoil force F produced by the emitted neutrino on a cantilever tip [1]:

$$F = \frac{\ln(2)NE_\nu S_\nu}{cT_{1/2}} \qquad (7)$$

Here $N$ is the number of radioactive atoms, $E_\nu$ is the energy of the emitted neutrino. It was estimated in [1] that for the mass of the radioactive antimony atoms $1.4 \times 10^{-10}$ kg ($6.9 \times 10^{14}$ atoms), a 1 pN recoil force can be achieved at a temperature of 25 mK. Our more accurate computation increases this value to 31 mK. This relaxes somewhat the requirements for the dilution refrigerator.

## IV. Conclusion

In conclusion, we have directly computed the spin excess for neutrino radiated by electron capture beta decay of radioactive antimony atoms assuming that $^{119}$Sb nuclei are partially polarized by the strong hyperfine field in a ferromagnetic sample. Our results support the idea that the radioactive antimony nuclei are the promising candidates for the compact source of directed mono-energetic neutrino radiation.

## Acknowledgment

The authors would like to sincerely thank L. M. Folan for very useful discussions.